# Limitations of Label-Free Sensors in Serum Based Molecular Diagnostics

*Manoj M. Varma[1,2]*

[1]*Center for Nano Science and Engineering, Indian Institute of Science, Bangalore*

[2]*Electrical Communication Engineering, Indian Institute of Science, Bangalore*

mvarma@cense.iisc.ernet.in

**Abstract**

Immunoassay formats applicable for clinical or point-of-care diagnostics fall into two broad classes. One which uses labeled secondary antibodies for signal transduction and the other which does not require the use of any labels. Comparison of the limits of detection (LoD) reported by these two sensing approaches over a wide range of detection techniques and target molecules in serum revealed that labeled techniques achieve 2-3 orders of magnitude better LoDs. Further, a vast majority of commercial tests and recent examples of technology translations are based on labeled assay formats. In light of this data, it is argued that extension of traditional labeled approaches and enhancing their functionality may have better clinical impact than the development of newer label-free techniques.

**Is a "mobile phone revolution" possible in personalized diagnostics?**

Ubiquitous access to Information and Communication Technology (ICT), particularly, wireless communication technologies, has radically transformed our lives with mobile-enabled technologies applied in areas ranging from education and entertainment to healthcare [1]. The deep penetration of mobile technologies, even in economically underdeveloped nations, is expected to provide better living conditions through mobile-enabled services in such countries. Analogous to the mobile revolution, we would like to see a radical transformation in human health monitoring enabled by ubiquitous access to affordable personal healthcare devices for continuous monitoring of human health parameters. The health parameters may include macroscopic state variables such as heart rate, ECG (Echocardiogram) and blood pressure as well as microscopic state variables such as the concentration of **biomarker** (text in bold is a glossary item) proteins in **serum**. One could use the example of ICT to imagine what the key enablers for such a transformation in personal health monitoring would be. Firstly, there should be sensing platforms or techniques where economies of scale can be applied to enable deep global penetration through cost reduction. In the case of ICT, Silicon based integrated electronics technology played this role [1]. In the case of personalized health monitoring it remains to be seen if such a universal platform technology would emerge. Secondly, not only the hardware cost, which is the capital expense, but also the recurring usage cost must be low. In the case of ICT this was possible due to the same economies of scale operating in the Silicon semiconductor manufacturing which led to low unit cost of data usage. In the case of healthcare devices, low usage cost translates to low unit cost of testing in terms of consumables required such as sensors and reagents. Finally, these devices must be easy to use with minimal user intervention required. This is necessary for the widespread adoption of such technology and to encourage its frequent use.

**Methods for diagnostic testing: labeled and label-free Immunoassays**

Currently health monitoring consists of a few simple tests which are possible at the clinician's office such as blood-pressure or ECG and a large number of tests requiring centralized testing labs causing delays of up to a few days between testing and results [2,3]. As a result, data from continuous monitoring of a given individual over his/her lifespan for parameters even as simple as blood-pressure are not easily available. Such datasets collected over a large number of individuals over time may reveal as yet undiscovered strategies for disease management or prevention. For several diseases including certain cancers, early detection exponentially improves the survival rate [4]. Even though one may not need day to day or even weekly testing frequency in these cases, there is certainly arguable merit in the development of sensing technologies which could perhaps be used at home and are capable of deep global penetration making early diagnosis accessible across economic strata.

Diagnostic tests are typically done using blood samples and are generally based on the **immunoassay** format where an antibody or a receptor is bound on a solid surface [5,6, Box 1]. These receptors bind the target biomarker from blood/serum or the sample under consideration. Ambitious goals such as early cancer detection requires the detection of biomarkers present in blood at extremely low concentrations with a high level of specificity using a system with few process steps for ease of use [7, 8]. Two broad classes of biosensors emerge depending on how the binding of target molecules is detected by the sensing instrument. The first class includes gold standards such as **ELISA** (Enzyme Linked Immunosorbent Assay) where a secondary antibody with a suitable label molecule conjugated to it, binds to the target-receptor complex immobilized on the surface [Box 1]. A better alternative, to enable ease of use, would be to detect the binding of the target molecules to the receptors directly without any secondary antibody labels. This approach is called label-free detection. By eliminating the need for labels and associated sample processing steps, label-free systems can potentially operate with minimal or no user intervention. By combining this simplicity of testing with the ability to detect small concentrations of biomarkers in complex samples such as serum, easy to use label-free sensors hold the promise of radically transforming personalized health diagnostics in a way comparable to the transformation of data communication with the advent of mobile phones.

**Box 1: Immunoassay format**
The solid phase immunoassay format developed during the 60's and 70's is a method to detect the presence of proteins or other molecules present in a sample using antibodies immobilized on a solid surface such as glass, nitrocellulose or Silicon [9]. As indicated in the figure, antibodies (blue Y shaped objects in the figure) immobilized on the solid surface capture the target molecules (red spheres) from the sample containing other molecules (blue spheres) which may potentially interfere by binding to the antibodies. This is called non-specific binding. The binding of the molecules (target or non-specific molecules) to the solid surface is prevented by appropriately chosen surface blocking steps so that the resultant signal from the sensor is entirely due to the interaction of the sample with the immobilized antibodies. Recently researchers have also developed other ways to capture target molecules, for example aptamers (specific DNA sequences), instead of antibodies. In general, the basic idea of immunoassay is that receptor molecules immobilized on a solid surface capture specific targets from a sample. In any given assay there will be some non-specific component of the signal which needs to be accounted for while interpreting the measurement. There are two broad classes of immunoassays depending on the detection of target receptor binding. One in which a secondary antibody conjugated with a label (blue Y shape with a green star attached to it) is used to bind to the target molecules captured by the immobilized antibodies making a sandwich structure. The signal is read out using properties of the label. For example, the label could be a **fluorophore** which emits light with a certain color or an enzyme molecule which converts a substrate to a colored product as in ELISA tests [10]. This type of detection is referred to as labeled detection. The labeling process and the incubation with secondary antibodies introduce additional process steps as well as increases the usage cost of such tests. As a result, several groups have developed techniques not requiring the use of labeled secondary antibodies. Instead, they may measure the optical refractive index, electrical conductance or mass change associated with the capture of target molecules. Such assays are called label-free assays as they do not use labeled secondary antibodies.

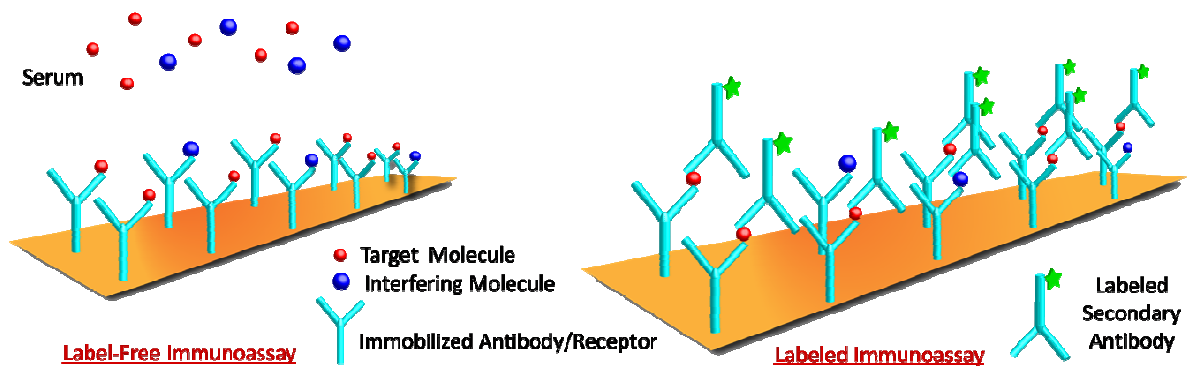

Figure 1 Label-Free and Labeled Immunoassay Schemes

**Comparison of limit of detection of labeled and label-free sensors**

Label-free sensors have witnessed decades of development with a veritable zoo of techniques available today exploiting physical effects as exotic as superconducting quantum interference [11]. However, in spite of this intense activity, most real-world tests including FDA approved tests are still based on labeled formats such as ELISA and immuno-fluorescence assays [12] and almost all recent examples from literature describing translation of diagnostic technology to real-world applications are also based on the labeled format [2,5-8,13-16]. It is therefore very important now to critically compare the performance of labeled and label-free technologies to understand if there are systematic reasons for the lack of prevalence for label-free technologies. To assess the current state of performance of label-free techniques, I compared the nearly 120 limit of detection (LoD) values reported for label-free and labeled methods for serum based biomarker detection. There were significant differences in the surface functionalization and assay protocols followed by different research groups reporting these **LoDs**. Consequently such a comparison may suffer from biases arising from the sensitivity of the LoD to the specific experimental protocols. To better compare these two approaches, I identified reports where labeled and label-free assays were performed simultaneously, eliminating any bias arising from differences in experimental protocols. Finally, I also examined recent examples describing translation of sensing techniques into real-world applications. It was found that labeled techniques significantly outperformed label-free techniques in all these contexts. In light of this observation, I argue that extension of traditional labeled assays into lab-on-chip formats and enhancing their performance using innovative signal read-out methods may have a better clinical impact than the frenzied development of newer label-free techniques we have witnessed in recent times.

The LoDs reported by various groups for a range of protein biomarkers, including those related to cancer were examined [17-76]. The complete data set along with the search strings used in the bibliographic database Web of Science [77] for literature survey is provided in Supplmentary Data Table [78]. Only articles reporting LoDs in serum were considered. For each category of data, namely, labeled and label-free, the cumulative distribution function (CDF) [Box 2] of the reported LoD values was constructed. A metric referred here as $LoD_{50}$ was defined based on the CDF. The $LoD_{50}$ is the LoD value for which the CDF function reaches the mid-point. The $LoD_{50}$ value therefore represents a kind of weighted average of the LoDs reported for each category. There were 54 data points for labeled detection and 53 data points for label-free detection making the comparison reasonable. Further, this comparison was done across proteins ranging from 10 kDa to 600 kDa with assays varying in the antibodies/receptors used, surface functionalization protocols and signal detection methods spanning mechanical, electrical, electrochemical and optical domains involving techniques as wide as Surface Plasmon Resonance, e.g. [41], micro-cantilevers, e.g. [51], fluorescence-immunoassays, e.g. [32], ELISA e.g. [54], optical interferometry, e.g. [25] and Silicon nanowires, e.g. [26]. Therefore, the data collected was a comprehensive representation of serum based labeled and label-free detection approaches. In order to extract the $LoD_{50}$ value in an objective manner, the data points were fitted with a smooth curve. As seen in Fig. 2 (a), the $LoD_{50}$ value of labeled detection was about 0.1 pM (pico-Molar) while that of label-free detection was around 10 pM indicating a 3 orders of magnitude gap in LoDs in favor of labeled detection methods. An examination of the high performing label-free detection techniques, depicted as "label-free outliers" in Fig. 2 a) revealed that all of them used significant amplification of the signal by using secondary antibodies tagged with micron sized beads, nanoparticles or enzymes. Use of such tagged secondary antibodies is contrary to the label-free detection paradigm and it is debatable whether these techniques should indeed be classified as label-free. Such reports were classified in a new category called "Label-free secondary amplified" and a reanalysis of the data was done. Out of the 53 data points in the label-free category, 30 of them, i.e. more than 50%, used secondary amplification. The CDF of the different sensing approaches classified into labeled, label-free and label-free secondary amplified is shown in Fig. 2 (b). Understandably, amplification of the label-free signal using secondary antibodies results in an order of magnitude improvement achieving an $LoD_{50}$ value around 3 pM compared to the $LoD_{50}$ of about 30 pM for direct label-free detection. However this performance is still nearly two orders of magnitude worse than that achieved by labeled detection which has a $LoD_{50}$ around 0.1 pM.

> **Box 2: Cumulative Distribution Function**
> Cumulative Distribution Function (*CDF*) is a function used in statistics to characterize the distribution of observed values of a variable [79]. *CDF* curves are generally normalized to a maximum of one so that the range of *CDF* is always from 0 to 1. The value of a normalized *CDF* curve for a variable *V* at *x*, denoted by $CDF_V(x)$, represents the probability that a random measurement of *V* would turn up a value less than *x*. For e.g. if the *CDF* for some variable *V* at *x* = 100 is 0.05, it means that it is very unlikely for measurements of *V* to yield values less than 100. We should expect the measurement of *V* to almost always yield values greater than 100. In the context of this article, *CDF(x)* represents the probability that a given research article in the respective detection category (labeled or label-free) would report a LoD less than *x*. For example, the value of *CDF* curve for label-free detection at LoD of 1 pM is about 0.25 (Figure 2 (a)) while it is about 0.9 for labeled detection. This means that if we picked a random research article dealing with label-free detection there is only a 25% chance that it would report a LoD of less than 1 pM while there is a 90% chance that the reported LoD would be less than 1 pM if the article was dealing with labeled detection. This is the basis for the observation that labeled techniques appear to outperform label-free methods. To construct the approximate *CDF* curve from a set of observed values, we rank the observations in ascending order and use $CDF(x) = \frac{R(x)-1}{N-1}$ where *R(x)* is the rank (position) of *x* in the sorted list of observations and *N* is the total number of observations in the set. We then plot *CDF(x)* against *x* for each observed value in the set to obtain the *CDF* curves shown in Fig. 2 (a) and (b).

To check if there was any correlation between reported LoDs with the molecular mass of the target biomarker, the data was re-plotted in the manner shown in Fig. 2 (c). The horizontal axis is the reported LoD in pM (pico-molar) while the vertical axis is the molecular weight of the target biomarker. Red, green and blue stripes represent labeled, label-free secondary amplified and direct label-free LoDs respectively. We do not see any correlation associated with the different targets considered. In other words the conclusions drawn from Figs. 2 (a) and (b) are valid across the entire range of target biomarkers considered. The conclusion emerging from this analysis is that labeled detection techniques are 2-3 orders of magnitude more sensitive than label-free approaches and it is only after significant signal amplification using tagged secondary antibodies that label-free approaches can attain similar performance.

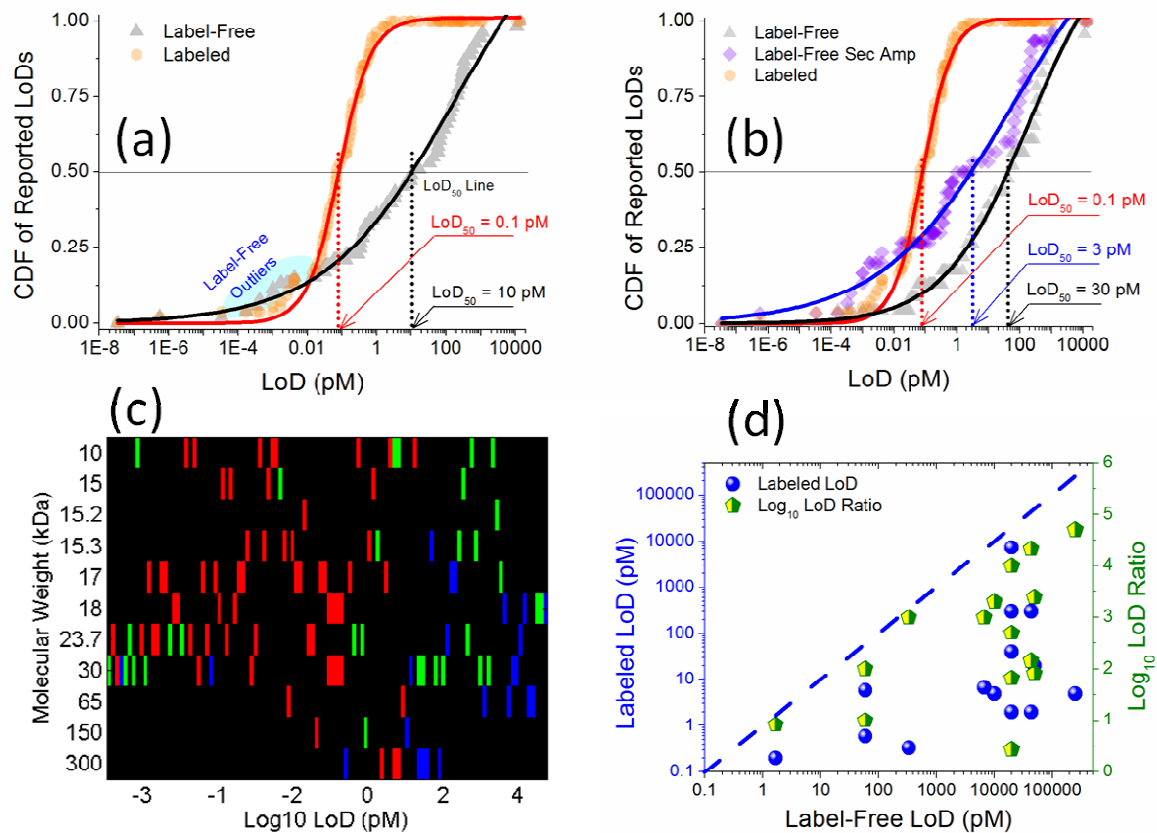

Figure 2 Analysis of LoDs reported in literature. Fig. 2 (a) shows the CDF [Box1] of labeled (red curve) and label-free (black curve) LoDs. The $LoD_{50}$, which represents a weighted average, of labeled detection is two orders of magnitude better than that of label-free detection. It was found that label-free techniques which significantly outperformed their peers (depicted as label-free outliers in Fig. 2 (a)) used tagged secondary antibodies for signal amplification. These LoD values were classified in a separate category called label-free secondary amplified ("Label-free Sec Amp" in Fig. 2 (b)) and the CDF curves for the 3 categories were plotted in Fig. 2 (b). Labeled techniques still outperformed label-free techniques in spite of secondary amplification by more than an order of magnitude based on the $LoD_{50}$ values shown in Fig. 2 (b). Fig. 2 (c) shows that there is no correlation between the achieved LoDs and the molecular weight of the targeted biomarker. Red, green and blue stripes represent labeled, label-free secondary amplified and direct label-free LoDs respectively. Fig. 2 (d) shows the comparison of direct label-free and labeled LoDs measured simultaneously in the same assay eliminating variations arising from differences in experimental protocols. It is seen that every data point shows better performance

of the labeled technique. The log10 ratio of the label-free to labeled LoD is plotted on the right vertical axis. The mean log10 LoD ratio is around 3 which supports the conclusion from Fig. 2 (a) and (b) that labeled techniques achieve 2-3 orders magnitude better LoDs compared to label-free techniques.

**Data from simultaneous labeled and label-free assays**

As mentioned earlier, LoD is sensitive to surface functionalization methods and associated experimental protocols. As there are likely to be major variations in these parameters for the data shown in Fig. 2 (a), the best way to compare the two different detection approaches would be to measure the same assay using a labeled and a label-free method simultaneously. Although few in number, some research groups have indeed done such experiments enabling direct comparison of labeled and label-free detection techniques without any confounding factors [24,33,61,72,73]. Fig. 2 (d) plots the label-free LoD on the horizontal axis against the LoD reported for corresponding labeled assay on the left vertical axis. On the right vertical axis the base-10 logarithm of the ratio of label-free LoD to the labeled LoD is plotted. For every single data point in Fig. 2 (d), it is seen that labeled LoD was significantly better than the corresponding label-free LoD. Moreover, the mean $\log_{10}$ ratio of label-free LoD to the labeled LoD of about 3 supports the conclusions drawn from the previous analysis, namely that labeled approaches achieve 2-3 orders of magnitude better LoDs than label-free techniques.

**Examination of outlier technologies**

It is interesting to examine the best performers in each category to identify techniques that show the highest promise. In the case of labeled detection, Plasmonic ELISA [35] and Digitial ELISA [54] report the best LoDs. In the case of label-free detection, inverse sensitivity assay employing enzyme mediated nanoparticle synthesis [34], massive signal amplification with 1 micron diameter magnetic beads [29], with enzyme conjugated magnetic beads [52] and so on report the best LoDs. However as pointed out earlier, this approach is only superficially different from using fluorophore or enzyme labels as done in labeled immunoassays and it is debatable if they should even be considered as label-free techniques. Direct label-free detection using Si nanowires has reported extremely low LoDs using desalted serum [36]. However, desalting may pose sample pre-treatment challenges. More importantly, such a dramatic performance advantage of nanowires over other label-free techniques such as SPR, micro-nano-mechanical resonators or even electrochemical methods must really be supported on firm theoretical grounds which is currently lacking. To summarize, the analysis of LoDs reported in serum strongly suggests that secondary antibody labels are required to achieve performance compared to traditional labeled techniques such as ELISA or fluorescence immunoassays. However, amplification using tagged secondary antibodies runs contrary to the label-free detection paradigm of direct detection with minimal sample processing and perhaps such approaches shouldn't even be considered as label-free.

**Why might label-free assays perform worse than labeled assays?**

Secondary antibody amplified label-free $LoD_{50}$ is still at least an order of magnitude worse than the $LoD_{50}$ of labeled approach even though both approaches use labels. Ignoring multiplicative factors, LoD can be written as $LoD = \sigma_{noise}/S$ where $\sigma_{noise}$ is the noise floor of measurement and $S$ is the **sensitivity** of the measurement technique [80]. It is believed that the noise floor in the current generation of biosensors mostly arises from non-specific binding (NSB) processes, for e.g., binding of interfering molecules to the receptors [81]. In this case, the noise floor will be the standard deviation of signal produced by a negative control sample, which in the case of serum based tests will be serum devoid of the target biomarker. In the case of labeled techniques, $\sigma_{noise}$ mainly arises from NSB of the secondary antibody to the target biomarker or unblocked sensor surface. However, in the case of label-free approaches, irrespective of whether they are amplified or not, $\sigma_{noise}$ can arise due to the NSB of the secondary antibody, NSB of the target biomarker, erosion of receptors [67] or similar phenomena related to the functionalization layers. In other words the noise floor of label-free detection is likely to be larger than that of labeled approaches. The improvement in $LoD_{50}$ for secondary antibody tagged amplified label-free sensors arises from increased sensitivity (larger $S$) due to signal amplification. This is a plausible model to explain the observations in Fig. 2.

**Conclusions and implications for clinical assay development**

Perhaps related to the observations made above, examination of commercial tests for FDA approved serum biomarkers [12] or the list of recent examples demonstrating the translation of biosensing techniques from lab to the real world [2,5-8,13-16], we see that almost all of them are based on labeled assay formats with color based or fluorescence readouts. It is impossible to find even a single example in this domain which uses a sophisticated label-free technique such as nanowires or nano-mechanical resonators. On the other hand many of these technology translations employ clever variations of the labeled detection strategies. In this regard, recent extensions of conventional labeled detection techniques such as Digital-ELISA [55], Plasmonic-ELISA [35], inverse sensitivity assay [34] and so on, show great promise. In light of these observations, I argue that extension of traditional labeled assays into lab-on-chip formats and enhancing their performance using innovative signal read-out methods may have a better clinical impact than the continued development of newer label-free techniques leading to a highly fragmented techno-commerical landscape unlike the dominant Silicon semiconductor platform technology which formed the basis for the digital revolution. Digital-ELISA and Plasmonic-ELISA are notable recent examples in this direction. However, this is not to suggest that label-free techniques serve no purpose. There are several applications involving detection of molecules in simple samples such as buffers, or those involving large molecules or markers that appear in large concentration, where label-free techniques may indeed leverage their simplicity and cost advantages. However, when the requirement is to measure ultra-low (sub pM) concentration of markers from a complex sample such as serum, it appears that labeled detection strategies currently have a significant performance advantage over label-free techniques. It is hoped that the data presented here will stimulate discussions leading to a critical and realistic assessment of the capabilities of label-free techniques and help identify applications where their unique strengths can be exploited.

**Glossary Items**

**Biomarkers:** Biomarkers are molecules whose concentration in serum could indicate a diseased condition or predict the imminent development of disease. For e.g. elevated serum concentration of a protein called Prostate Specific Antigen (PSA) could indicate the presence of prostate cancer.

**ELISA**: Enzyme Linked Immunosorbent Assay (ELISA) is a commonly used immunoassay format where an enzyme labelled secondary antibody is used to bind targets captured by primary antibodies immobilized on a solid surface. Signal detection is based on the color change produced by the action of the enzyme on molecules referred to as substrate.

**Fluorophore**: A fluorophore is a molecule which can absorb light around a peak absorption wavelength and emit light at slightly longer wavelengths. They can be used for labelling secondary antibodies in labelled immunoassays.

**Immunoassay**: Immunoassay is the term used to describe the method to test for the presence of a molecule of interest, referred to as the target, using another molecule having a specific affinity for the target, such as an antibody against the target.

**LoD**: Limit of Detection (LoD) is the smallest concentration of targeted molecule that can be detected by a sensing technique. It is directly proportional to the ratio of noise floor of measurement and the sensitivity.

**Sensitivity**: Sensitivity of a sensor is the slope of the signal response to the stimulus. In the case of biosensors, it is the signal response per unit change in the concentration of the target molecule.

**Serum**: Serum is the fraction of blood without the cells and clotting factor. It contains several potential biomarkers.

This is the supplementary data table for the article **Limitations of Label-Free Sensors in Serum Based Molecular Diagnostics; Manoj M. Varma.** The table below shows the limit of detection (LoD) collected for several label-free (LF) and labeled immunoassay methods. Label-free detection is further classified into LF direct and LF sec amp. "LF sec amp" refers to the use of tagged secondary antibodies for signal amplification. The reference numbers are provided as they occur in the main text. The search strings used for collecting this data are given in the next sheet.

| Molecule | Molecular weight (Mw) in Literature | Mw Assumed (kDa) | Detection Category | Technique | Sample Type | LoD (From Ref) | LoD (ng/mL) | LoD (pM) | Reference |
|---|---|---|---|---|---|---|---|---|---|
| PSA | ~ 30 kDa | 30 | Labeled | FL (Fluorescence) | goat serum | 1 fg/mL | 0.000001 | 3.33E-05 | 17 |
| PSA | ~ 30 kDa | 30 | Labeled | SPR (Surface Plasmon Resonance) enhanced FL | female human plasma | 2 pg/mL | 0.002 | 6.67E-02 | 18 |
| IL-1 (a/b) | 17 kDa | 17 | LF direct | Fiber Optic SPR | 4% bovine serum | 1 ng/mL | 1 | 5.88E+01 | 19 |
| IL-6 | 23.7 kDa | 23.7 | LF direct | Fiber Optic SPR | CCM (Cell Culture Media) with 4% FBS (Fetal Bovine Serum) | 1 ng/mL | 1 | 4.22E+01 | 19 |
| TNF-a | ~ 17 kDa | 17 | LF direct | Fiber Optic SPR | CCM with 10% FBS | 1 ng/mL | 1 | 5.88E+01 | 19 |
| IL-6 | 23.7 kDa | 23.7 | Labeled | Luminex | pooled human plasma | 0.75 pg/mL | 0.00075 | 3.16E-02 | 20 |
| IL-6 | 23.7 kDa | 23.7 | Labeled | ELISA | pooled human plasma | 70 fg/mL | 0.00007 | 2.95E-03 | 20 |
| IL-8 | ~ 10 kDa | 10 | Labeled | Luminex (Bead FL) | pooled human plasma | 1.3 pg/mL | 0.0013 | 1.30E-01 | 20 |
| IL-8 | ~ 10 kDa | 10 | Labeled | ELISA | pooled human plasma | 0.26 pg/mL | 0.00026 | 2.60E-02 | 20 |

| Analyte | MW | MW (kDa) | Method | Platform | Matrix | LOD | LOD (ng/mL) | Ratio | Ref |
|---|---|---|---|---|---|---|---|---|---|
| PSA | ~ 30 kDa | 30 | LF direct | Si NW | desalted donkey serum | 1 pg/mL | 0.001 | 3.33E-02 | 21 |
| PSA | ~ 30 kDa | 30 | LF sec amp | Si cantilever | fetal bovine serum | 50 fg/mL | 0.00005 | 1.67E-03 | 22 |
| IFN-g | 17 - 19 KDa | 18 | Labeled | MSD (Meso Scale Discovery) | spiked human plasma | 0.7 pg/mL | 0.0007 | 3.89E-02 | 23 |
| IL-2 | 15.3 kDa | 15.3 | Labeled | Luminex | spiked human plasma | 8.7 pg/mL | 0.0087 | 5.69E-01 | 23 |
| IL-2 | 15.3 kDa | 15.3 | Labeled | MSD | spiked human plasma | 2.5 pg/mL | 0.0025 | 1.63E-01 | 23 |
| IL-4 | 15 kDa | 15 | Labeled | Luminex | spiked human plasma | 8.6 pg/mL | 0.0086 | 5.73E-01 | 23 |
| IL-4 | 15 kDa | 15 | Labeled | MSD | spiked human plasma | 0.7 pg/mL | 0.0007 | 4.67E-02 | 23 |
| IL-8 | ~ 10 kDa | 10 | Labeled | Luminex | spiked human plasma | 7.7 pg/mL | 0.0077 | 7.70E-01 | 23 |
| IL-8 | ~ 10 kDa | 10 | Labeled | MSD | spiked human plasma | 0.7 pg/mL | 0.0007 | 7.00E-02 | 23 |
| TNF-a | ~ 17 kDa | 17 | Labeled | Luminex | spiked human plasma | 6 pg/mL | 0.006 | 3.53E-01 | 23 |
| TNF-a | ~ 17 kDa | 17 | Labeled | MSD | spiked human plasma | 3 pg/mL | 0.003 | 1.76E-01 | 23 |
| PSA | ~ 30 kDa | 30 | Labeled | SPR enh FL | human serum | 10 pg/mL | 0.01 | 3.33E-01 | 24 |
| PSA | ~ 30 kDa | 30 | LF direct | SPR | human serum | 10 ng/mL | 10 | 3.33E+02 | 24 |
| PSA | ~ 30 kDa | 30 | LF sec amp | Reflectance | human serum | 4 ng/mL | 4 | 1.33E+02 | 25 |
| PSA | ~ 30 kDa | 30 | LF direct | Si nanowire | native human serum | 10 ng/mL | 10 | 3.33E+02 | 26 |
| IL-1 (a/b) | 17 kDa | 17 | Labeled | Luminex, MSD etc | human serum | ~ 1pg/mL | 0.001 | 5.88E-02 | 27 |
| IL-2 | 15.3 kDa | 15.3 | Labeled | Luminex | spiked human serum | 2.1 pg/mL | 0.0021 | 1.37E-01 | 27 |
| IL-6 | 23.7 kDa | 23.7 | Labeled | Luminex | human serum | 0.6 pg/mL | 0.0006 | 2.53E-02 | 27 |
| PSA | ~ 30 kDa | 30 | LF sec amp | QCM (Quartz Crystal Microbalance) | 75% human serum | 300 pg/mL | 0.3 | 1.00E+01 | 28 |

| Analyte | MW | MW (kDa) | Format | Method | Matrix | LOD | | | Ref |
|---|---|---|---|---|---|---|---|---|---|
| PSA | ~ 30 kDa | 30 | LF sec amp | SPR | calf serum | 10 fg/mL | 0.00001 | 3.33E-04 | 29 |
| PSA | ~ 30 kDa | 30 | LF sec amp | SPR | calf serum | 10 pg/mL | 0.01 | 3.33E-01 | 29 |
| IL-1 (a/b) | 17 kDa | 17 | Labeled | FL | human serum | 12 pg/mL | 0.012 | 7.06E-01 | 30 |
| IL-6 | 23.7 kDa | 23.7 | Labeled | FL | spiked human serum | 4.3 pg/mL | 0.0043 | 1.81E-01 | 30 |
| TNF-a | ~ 17 kDa | 17 | Labeled | FL | spiked human serum | 5 pg/mL | 0.005 | 2.94E-01 | 30 |
| IL-6 | 23.7 kDa | 23.7 | Labeled | Luminex | human serum | 1.6 pg/mL | 0.0016 | 6.75E-02 | 31 |
| IL-6 | 23.7 kDa | 23.7 | Labeled | MSD | human serum | 1 pg/mL | 0.001 | 4.22E-02 | 31 |
| IL-8 | ~ 10 kDa | 10 | Labeled | Luminex | human serum | 4 pg/mL | 0.004 | 4.00E-01 | 31 |
| IL-8 | ~ 10 kDa | 10 | Labeled | MSD | human serum | 0.2 pg/mL | 0.0002 | 2.00E-02 | 31 |
| IFN-g | 17 - 19 KDa | 18 | Labeled | Luminex | human Serum | 1 pg/mL | 0.001 | 5.56E-02 | 32 |
| IL-2 | 15.3 kDa | 15.3 | Labeled | Luminex | spiked human serum | 1.1 pg/mL | 0.0011 | 7.19E-02 | 32 |
| IL-4 | 15 kDa | 15 | Labeled | Luminex | spiked human serum | 0.6 pg/mL | 0.0006 | 4.00E-02 | 32 |
| IL-5 | 15.2 kDa | 15.2 | Labeled | Luminex | spiked human serum | 2.7 pg/mL | 0.0027 | 1.78E-01 | 32 |
| IL-6 | 23.7 kDa | 23.7 | Labeled | Luminex | spiked human serum | 4.6 pg/mL | 0.0046 | 1.94E-01 | 32 |
| IL-8 | ~ 10 kDa | 10 | Labeled | Luminex | spiked human serum | 1 pg/mL | 0.001 | 1.00E-01 | 32 |
| TNF-a | ~ 17 kDa | 17 | Labeled | Luminex | spiked human serum | 600 fg/mL | 0.0006 | 3.53E-02 | 32 |
| TNF-a | ~ 17 kDa | 17 | LF direct | SPR | buffer | 1 ng/mL | 1 | 5.88E+01 | 33 |
| TNF-a | ~ 17 kDa | 17 | Labeled | SP coupled FL | buffer | 10 pg/mL | 0.01 | 5.88E-01 | 33 |
| TNF-a | ~ 17 kDa | 17 | Labeled | ELISA | buffer | 100 pg/mL | 0.1 | 5.88E+00 | 33 |
| PSA | ~ 30 kDa | 30 | LF sec amp | LSPR | human serum | 1 attog/mL | 1E-09 | 3.33E-08 | 34 |
| PSA | ~ 30 kDa | 30 | LF sec amp | LSPR | human serum | 1 attog/mL | 1E-09 | 3.33E-08 | 35 |

| | | | | | | L | | | |
|---|---|---|---|---|---|---|---|---|---|
| PSA | ~ 30 kDa | 30 | LF direct | Si nanowire | human serum | 5 fg/mL (dubious) | 0.000005 | 1.67E-04 | 36 |
| PSA | ~ 30 kDa | 30 | Labeled | FL | human serum | 0.8 pg/mL | 0.0008 | 2.67E-02 | 37 |
| PSA | ~ 30 kDa | 30 | LF sec amp | SPR | 20% human serum | 5 ng/mL | 5 | 1.67E+02 | 38 |
| ALCAM | 65 kDa | 65 | LF direct | SPR | 10% human serum | 20 ng/mL | 20 | 3.08E+02 | 39 |
| ALCAM | 65 kDa | 65 | Labeled | ELISA | human serum | 0.1 ng/mL | 0.1 | 1.54E+00 | 39 |
| ALCAM | 65 kDa | 65 | LF direct | SPR | human serum | 64 ng/mL | 64 | 9.85E+02 | 40 |
| ALCAM | 65 kDa | 65 | LF direct | imaging SPR | 10% human plasma | 45 ng/mL | 45 | 6.92E+02 | 41 |
| ALCAM | 65 kDa | 65 | LF direct | Suspended microchannel resonator (SMR) | undiluted fetal bovine serum | 10 ng/mL | 10 | 1.54E+02 | 42 |
| CA125 | 0.2 - 1 MDa | 600 | LF direct | QCM | human serum/plasma | 5 U/mL | 5 | 8.33E+00 | 43 |
| CA125 | 0.2 - 1 MDa | 600 | LF direct | Electrical impedance | human plasma | 1 U/mL | 1 | 1.67E+00 | 44 |
| CA125 | 0.2 - 1 MDa | 600 | LF direct | Electrochemical | human serum | 0.1 U/mL | 0.1 | 1.67E-01 | 45 |
| CA125 | 0.2 - 1 MDa | 600 | LF direct | Optical cavity | human serum | 1.8 U/mL | 1.8 | 3.00E+00 | 46 |
| IL-2 | 15.3 kDa | 15.3 | Labeled | Electrochemiluminescence | human serum | 1 pg/mL | 0.001 | 6.54E-02 | 47 |
| IL-4 | 15 kDa | 15 | LF sec amp | Electrochemilum | human serum | 2 pg/mL | 0.002 | 1.33E-01 | 47 |
| TNF-a | ~ 17 kDa | 17 | Labeled | Bead FL | human serum | 3 pg/mL | 0.003 | 1.76E-01 | 48 |
| TNF-a | ~ 17 kDa | 17 | Labeled | ELISA | human serum | 5 pg/mL | 0.005 | 2.94E-01 | 48 |
| IL-2 | 15.3 kDa | 15.3 | LF sec amp | MRR (micro-ring resonator) | CCM with 10% FBS | 1.9 ng/mL | 1.9 | 1.24E+02 | 49 |
| IL-4 | 15 kDa | 15 | LF sec amp | MRR | CCM with 10% FBS | 1 ng/mL | 1 | 6.67E+01 | 49 |
| IL-5 | 15.2 kDa | 15.2 | LF sec amp | MRR | CCM with 10% FBS | 3.4 ng/mL | 3.4 | 2.24E+02 | 49 |

| Analyte | MW | MW (kDa) | Format | Method | Matrix | LOD | LOD (ng/mL) | LOD (pM) | Ref |
|---|---|---|---|---|---|---|---|---|---|
| TNF-a | ~ 17 kDa | 17 | LF sec amp | MRR | CCM with 10% FBS | 4.6 ng/mL | 4.6 | 2.71E+02 | 49 |
| IL-8 | ~ 10 kDa | 10 | Labeled | FL | human serum | 26 pg/mL | 0.026 | 2.60E+00 | 50 |
| IL-6 | 23.7 kDa | 23.7 | LF sec amp | Si cantilever | human serum | 100 fg/mL | 0.0001 | 4.22E-03 | 51 |
| IL-6 | 23.7 kDa | 23.7 | LF sec amp | Electrochemical | calf serum | 10 fg/mL | 0.00001 | 4.22E-04 | 52 |
| IL-8 | ~ 10 kDa | 10 | LF sec amp | Electrochemical | calf serum | 10 fg/mL | 0.00001 | 1.00E-03 | 52 |
| IL-2 | 15.3 kDa | 15.3 | LF sec amp | MRR | CCM with 10% FBS | ~ 10 pg/mL | 0.01 | 6.54E-01 | 53 |
| IL-6 | 23.7 kDa | 23.7 | LF sec amp | MRR | CCM with 10% FBS | ~ 10 pg/mL | 0.01 | 4.22E-01 | 53 |
| IL-8 | ~ 10 kDa | 10 | LF sec amp | MRR | CCM with 10% FBS | ~ 10 pg/mL (check) | 0.01 | 1.00E+00 | 53 |
| IL-1 (a/b) | 17 kDa | 17 | Labeled | ELISA | bovine serum | 30 fg/mL | 0.00003 | 1.76E-03 | 54 |
| IL-6 | 23.7 kDa | 23.7 | Labeled | Digital ELISA | spiked bovine serum | 24 fg/mL | 0.000024 | 1.01E-03 | 54 |
| TNF-a | ~ 17 kDa | 17 | Labeled | Digital ELISA | human serum | 70 fg/mL | 0.00007 | 4.12E-03 | 54 |
| IFN-g | 17 - 19 KDa | 18 | Labeled | FL | human serum | 6 pg/mL | 0.006 | 3.33E-01 | 55 |
| IFN-g | 17 - 19 KDa | 18 | Labeled | Plas Enh NIR FL | 10% fetal bovine serum | 250 fg/mL | 0.00025 | 1.39E-02 | 55 |
| IL-1 (a/b) | 17 kDa | 17 | Labeled | Plas Enh NIR FL | 10% fetal bovne serum | 70 fg/mL | 0.00007 | 4.12E-03 | 55 |
| IL-4 | 15 kDa | 15 | Labeled | Plas Enh NIR FL | 10% fetal bovne serum | 1.3 pg/mL | 0.0013 | 8.67E-02 | 55 |
| IL-6 | 23.7 kDa | 23.7 | Labeled | Plas Enh NIR FL | 10% fetal bovne serum | 60 fg/mL | 0.00006 | 2.53E-03 | 55 |
| TNF-a | ~ 17 kDa | 17 | Labeled | Plas Enh NIR FL | 10% fetal bovne serum | 470 fg/mL | 0.00047 | 2.76E-02 | 55 |
| IFN-g | 17 - 19 KDa | 18 | LF direct | SPR | 100x diluted bovine serum | 250 ng/mL | 250 | 1.39E+04 | 56 |
| IFN-g | 17 - 19 KDa | 18 | LF direct | Electrochemical | RPMI/10% fetal bovine serum | 5 ng/mL | 5 | 2.78E+02 | 57 |

| Analyte | MW | MW (kDa) | Format | Detection | Matrix | LOD | LOD (ng/mL) | LOD (nM) | Ref |
|---|---|---|---|---|---|---|---|---|---|
| IFN-g | 17 - 19 KDa | 18 | LF direct | SPR | 2% depleted human plasma | 10 ng/mL | 10 | 5.56E+02 | 58 |
| IFN-g | 17 - 19 KDa | 18 | LF sec amp | Electrochemical | fetal bovine serum | 200 ng/mL | 200 | 1.11E+04 | 59 |
| IFN-g | 17 - 19 KDa | 18 | LF direct | Chemiluminescence | 5% human serum | 20 ng/mL | 20 | 1.11E+03 | 60 |
| Rabbit IgG | 150 kDa | 150 | Labeled | BioCD (Interferometry) | bovine serum | 30 pg/mL | 0.03 | 2.00E-01 | 61 |
| Rabbit IgG | 150 kDa | 150 | LF direct | BioCD | bovine serum | 250 pg/mL | 0.25 | 1.67E+00 | 61 |
| Rabbit IgG | 150 kDa | 150 | LF sec amp | BioCD | bovine serum | 70 pg/mL | 0.07 | 4.67E-01 | 61 |
| IL-2 | 15.3 kDa | 15.3 | LF Direct | BSI (Backscattering Interferometry) | media containing 1% serum | 150 pg/mL | 0.15 | 9.80E+00 | 62 |
| IL-2 | 15.3 kDa | 15.3 | LF sec amp | MRR | CCM with 10% FBS | 1 ng/mL | 1 | 6.54E+01 | 63 |
| IL-8 | ~ 10 kDa | 10 | LF sec amp | MRR | CCM with 10% FBS | 1 ng/mL | 1 | 1.00E+02 | 63 |
| IL-6 | 23.7 kDa | 23.7 | LF sec amp | Electrochemical | serum | 410 fg/mL | 0.00041 | 1.73E-02 | 64 |
| IL-6 | 23.7 kDa | 23.7 | LF sec amp | SPR | CCM with 10% FBS | 2 ng/mL | 2 | 8.44E+01 | 65 |
| IL-6 | 23.7 kDa | 23.7 | LF direct | SPR | CCM with 10% FBS | > 12 ng/mL | 12 | 5.06E+02 | 65 |
| IL-6 | 23.7 kDa | 23.7 | LF sec amp | GMR (Giant Magneto-resistance) | serum | 9 pg/mL | 0.009 | 3.80E-01 | 66 |
| IL-6 | 23.7 kDa | 23.7 | LF direct | Reflectance | buffer | 19 ng/mL | 19 | 8.02E+02 | 67 |
| IL-6 | 23.7 kDa | 23.7 | LF sec amp | Reflectance | buffer | 2 ng/mL | 2 | 8.44E+01 | 67 |
| IL-6 | 23.7 kDa | 23.7 | Labeled | Nano Enh FL | buffer and serum | 12.9 attog/mL | 1.3E-08 | 5.49E-07 | 68 |
| IL-8 | ~ 10 kDa | 10 | LF sec amp | SPR | human saliva | 2.2 ng/mL | 2.2 | 2.20E+02 | 69 |
| PSA | ~ 30 kDa | 30 | LF sec amp | Si cantilever | human serum | 30 fg/mL | 0.00003 | 1.00E-03 | 70 |

| Analyte | MW | MW (kDa) | Format | Method | Matrix | LOD | LOD (ng/mL) | LOD (pM) | Ref |
|---|---|---|---|---|---|---|---|---|---|
| PSA | ~ 30 kDa | 30 | LF sec amp | Electrochemical | 100% human serum | 1 ng/mL | 1 | 3.33E+01 | 71 |
| PSA | ~ 30 kDa | 30 | LF sec amp | Electrochemical | 10% human serum | 0.1 ng/mL | 0.1 | 3.33E+00 | 71 |
| ssDNA (25bp) | 8.2 kDa | 8.2 | LF direct | imaging SPR | Buffer | 10 nM | | 1.00E+04 | 72 |
| ssDNA (25bp) | 8.2 kDa | 8.2 | LF direct | QCM | Buffer | 250 nM | | 2.50E+05 | 72 |
| ssDNA (25bp) | 8.2 kDa | 8.2 | Labeled | FL | Buffer | 1 - 10 pM | | 5.00E+00 | 72 |
| Estradiol | 272 Da | 0.272 | Labeled | ELISA | Buffer; 4 diff Abs in 3 diff sensing methods | 7.3 nM | | 7.30E+03 | 73 |
| Estradiol | 272 Da | 0.272 | LF direct | SPR | Buffer | 20 nM | | 2.00E+04 | 73 |
| Estradiol | 272 Da | 0.272 | Labeled | FL | Buffer | 40 pM | | 4.00E+01 | 73 |
| Estradiol | 272 Da | 0.272 | Labeled | ELISA | Buffer | 0.3 nM | | 3.00E+02 | 73 |
| Estradiol | 272 Da | 0.272 | LF direct | SPR | Buffer | 20 nM | | 2.00E+04 | 73 |
| Estradiol | 272 Da | 0.272 | Labeled | FL | Buffer | 2 pico M | | 2.00E+00 | 73 |
| Estradiol | 272 Da | 0.272 | Labeled | ELISA | Buffer | 0.6 nM | | 6.00E+02 | 73 |
| Estradiol | 272 Da | 0.272 | LF direct | SPR | Buffer | 49 nM | | 4.90E+04 | 73 |
| Estradiol | 272 Da | 0.272 | Labeled | FL | Buffer | 20 pM | | 2.00E+01 | 73 |
| Estradiol | 272 Da | 0.272 | Labeled | ELISA | Buffer | 0.3 nM | | 3.00E+02 | 73 |
| Estradiol | 272 Da | 0.272 | LF direct | SPR | Buffer | 4.3 nM | | 4.30E+03 | 73 |
| Estradiol | 272 Da | 0.272 | Labeled | FL | Buffer | 2 pico M | | 2.00E+00 | 73 |
| ALCAM | 65 kDa | 65 | Labeled | ELISA | diluted serum | 0.03 ng/mL | 0.03 | 4.62E-01 | 74 |
| CA125 | 0.2 - 1 MDa | 600 | Labeled | Luminex | human serum | 2.7 pg/mL | 0.02 | 3.33E-02 | 75 |

| PSA | ~ 30 kDa | 30 | Labeled | ELISA | human serum | 10 pg/mL | 0.01 | 3.33E-01 | 76 |

Search Strings Used

| Technique | Technology Variant | Search String |
|---|---|---|
| Labeled | Fluorescence/ELISA | (ts = sens* OR ts = assay) AND (ti = fluor* OR ti = ELISA) AND (ts = analyte) |
| Label-free (LF) generic | All | (ts = sens* OR ts = assay) AND (ts = "label free" OR ts = label-free) AND (ts = analyte) |
| Limit of detection | All | (ts = "limit of detection") AND (ts = analyte) |
| LF Optical | SPR | (ts = sens* OR ts = assay) AND (ts = SPR OR ts = plasmon*) AND (ts = analyte) |
| | Optical cavities, Micro-ring resonators | (ts = sens* OR ts = assay) AND (ts = silicon OR ts = resona*) AND (ts = analyte) |
| | Interferometry | (ts = sens* OR ts = assay) AND (ts = interfer*) AND (ts = analyte) |
| | General Optical | (ts = sens* OR ts = assay) AND (ts = optic*) AND (ts = analyte) |
| | Reflectance | (ts = sens* OR ts = assay) AND (ts = reflect*) AND (ts = analyte) |
| LF Electrical | Nanowire | (ts = sens* OR ts = assay) AND (ts = silicon OR ts = nanowire OR ts = nano-wire OR ts = CNT OR ts = nano-tube) AND (ts = analyte) |
| | Impedance | (ts = sens* OR ts = assay) AND (ts = imped*) AND (ts = analyte) |
| LF Mass/mech | cantilever | (ts = sens* OR ts = assay) AND (ts = cantilever OR ts = MEMS OR ts = silicon OR ts = micro-mech* OR ts = micromech* OR ts = resona*) AND (ts = analyte) |
| | QCM | (ts = sens* OR ts = assay) AND (ts = QCM OR ts = "quartz crystal") AND (ts = analyte) |
| LF Electrochem | Electrochemical | (ts = sens* OR ts = assay) AND (ts = electroche*) AND (ts = analyte) |
| LF magnetic | | (ts = sens* OR ts = assay) AND (ts = *magnet*) AND (ts = analyte) |
| | | |
| | | The term "analyte" in the search strings above was substituted with the appropriate biomarker, e.g. PSA, IL6 and so on for each category listed above |